*Gigahertz Single-cavity Dual-comb Laser for Rapid Time-domain Spectroscopy: from Few Terahertz to Optical Frequencies*

Benjamin Willenberg[1,*,x], Christopher R. Phillips[1,*], Justinas Pupeikis[1], Sandro L. Camenzind[1], Lars Liebermeister[2], Robert B. Kohlhass[2], Björn Globisch[2], and Ursula Keller[1]

[1] Department of Physics, Institute for Quantum Electronics, ETH Zurich, 8093 Zurich, Switzerland
[2] Fraunhofer Institute for Telecommunications, Heinrich Hertz Institute, HHI, 10587 Berlin, Germany
* These authors contributed equally. x Corresponding author: bwillenb@phys.ethz.ch

**Abstract:** Dual-comb generation from a single laser cavity provides a simple and high-performance solution to time sampling applications. We demonstrate a compact single-cavity dual-comb laser operating at gigahertz repetition rates and high repetition rate differences up to more than 100 kHz with sub-100 fs pulse duration. The single cavity approach leads to passive common noise suppression resulting in ultra-low relative timing jitter and fully resolvable comb lines in free-running operation. We showcase the laser performance with two application demonstrations: (a) time-domain spectroscopy of acetylene in the near-infrared via computational comb line tracking and (b) free-space THz time domain spectroscopy and thickness-measurements via adaptive sampling. For (b) we use efficient state-of the art iron-doped InGaAs photoconductive antennas to generate and detect the THz light. Here we operate these devices with an efficient Yb-based gigahertz repetition rate laser for the first time. One optical comb generates the THz light, while the other probes it via equivalent time sampling. We obtain signal strengths comparable to reference measurements with MHz repetition rate Er-based laser systems while achieving close to 1 GHz spectral resolution (defined by the comb line spacing) and generating THz frequencies up to 3 THz. By carrying out a careful investigation of the noise properties of the laser we confirm that the free-running gigahertz dual-comb oscillator provides a rapid yet highly precise optical delay sweep from a simple setup. Therefore, our approach will be beneficial for high-update rate time sampling and time-domain spectroscopy applications.

*Introduction*

Femtosecond modelocked laser-based optical frequency combs [1–3] have enabled many metrology applications such as spectroscopy and precision ranging [4,5]. Dual optical frequency combs [6,7] are an interesting extension of the optical frequency comb involving a pair of combs with a slight but well defined difference in their pulse repetition rate. They enable comb line resolved heterodyne measurements because beating between each pair of comb lines yields a corresponding line in the easily accessible radio frequency domain. Dual comb sources are also powerful tools for the measurement technique of equivalent time sampling (ETS) [8], sometimes referred to as asynchronous optical sampling (ASOPS) [9], that uses the delay sweep between the two pulse trains. In this technique, a real time window of duration $1/f_{rep}$ is transferred into an equivalent time window of duration $1/\Delta f_{rep}$, where $f_{rep}$ is the repetition rate of one of the combs and $\Delta f_{rep}$ is the repetition rate difference between the two combs. This corresponds to a scaling of the time axis by the factor $f_{rep}/\Delta f_{rep}$. Because this method of delay scanning does not require any moving parts, it is possible to obtain much faster and longer-range scans compared to conventional mechanical delay line based pump-probe measurements. High update rates are an important performance frontier since they enable real-time material inspection and label-free imaging.

A critical parameter for optical frequency comb-based sensing techniques is the accessible wavelength range of the source. Many strong spectroscopic features lie outside the near-infrared wavelength range, implying that well-established laser technology operating in this wavelength range must be combined with frequency conversion schemes. For example, recent work has accessed the functional group region (3 to 5 μm) and the molecular fingerprint region (5 to 20 μm) via difference frequency generation [10], optical parametric oscillation [11,12], and optical rectification [13]. A special case of optical rectification is the generation of terahertz radiation (0.1 to 10 THz), which has seen a lot of attention in recent years due to the progress in efficient photoconductive antennas [14].

The THz range is of special interest for scientific and industrial applications since it allows for non-invasive detection and analysis of many materials that are opaque in the visible and infrared [15]. Applications include detection of spectroscopic features in the 1 to 5 THz range to distinguish between plastics and explosives that look visibly identical [16], quality control monitoring through opaque packaging, non-invasive layer thickness measurements of paint with µm-accuracies [17], high-resolution gas spectroscopy, and alternatives to x-ray technology for label-free analysis of biological tissue (since THz radiation is non-ionizing) [18]. These applications are commonly addressed by the terahertz time-domain spectroscopy (THz-TDS) technique.

In THz-TDS, one optical pulse train generates a train of single-cycle THz-pulses on an emitter device, while the other optical pulse train is delayed and equivalent time samples the THz field on a receiver device [19]. The progress in photoconductive antennas (PCAs) in the past decade has made them the preferred choice for table-top systems with conversion efficiencies as high as 3.4% in power [20] at moderate optical pulse energies of a few hundred picojoules. As well as PCA based experiments, terahertz generation using nonlinear crystals and ≫nJ level optical pulse energies has also received a great deal of attention [21,22].

Many PCA systems use lasers with repetition rates of around 100 MHz in combination with a mechanical delay stage to implement the equivalent time sampling of the THz waveform, but this imposes a severe trade-off between the speed and range of the scan. The same type of lasers can implement THz-TDS via ETS, but the corresponding long delay range of 10 ns is only needed for specific applications (such as measuring targets with long response times or the sharp absorption lines of molecular gases at low pressures) [15]. For many applications a shorter range (<1 ns) and a corresponding spectral resolution (>1 GHz) is sufficient, with examples including gas spectroscopy at ambient pressures, or detection of small variations in the thickness of thin-film layers [23]. Limiting the scan to a shorter range avoids the dead time at the end of the time window, which improves the signal-to-noise ratio since the signal of interest will occupy a larger fraction of the measurement window. Electronically controlled optical sampling (ECOPS) [24] and other techniques [25,26] have been developed to address this by electronically controlling the pulse-to-pulse delay over a limited range much smaller than $1/f_{rep}$. An alternative and potentially much simpler approach is to use a higher repetition rate free-running dual-comb laser. Gigahertz repetition rates enable high (multi-kHz) update rates while scanning the full delay range with ≪100 fs resolution. In the context of THz-TDS with PCAs, such lasers are also a promising path to boost the signal strength by using a higher average power while staying below the pulse energy damage threshold of the devices. Gigahertz lasers have been explored for pump-probe spectroscopy via Ti:sapphire lasers at 1 GHz [27] and 10 GHz [28], but the high cost of Ti:sapphire technology has hindered wider adoption.

Use of gigahertz lasers for dual-comb spectroscopy and THz-TDS has seen renewed interest in recent years due to advances in high repetition rate Ytterbium and Erbium based frequency combs [29–34]. Diode-pumped solid-state lasers with low-loss, low-nonlinearity, low-dispersion cavities are ideally suited for generating gigahertz combs [35,36], and they are far simpler than traditional Ti:sapphire systems while offering more damping of high-frequency pump intensity fluctuations. They also support lower noise [31], higher power, and exhibit more straightforward repetition rate scaling compared to fiber lasers.

Another critical consideration for practical deployment of dual-comb applications is system complexity. Conventional systems, consisting of a pair of locked femtosecond oscillators, are complex and require several feedback loops. A promising alternative are single-cavity dual-comb oscillators, where high coherence between the combs is achieved in free-running operation by having both combs share the same laser cavity. This approach has been demonstrated with semiconductor disk lasers [37], free-space bidirectional ring lasers [38], and bidirectionally modelocked fiber lasers [39], amongst others. Recently, we have demonstrated a set of free-running solid-state single-cavity systems with all common optics and ultra-low relative timing noise performance utilizing birefringent multiplexing [40–42] or spatial multiplexing [43,44]. The system reported in [43] allowed for sub-cycle relative timing jitter ([20 Hz, 100 kHz] integration range) and thereby surpasses the performance of ASOPS systems with two locked lasers for pump-probe measurements.

An important factor in obtaining this performance from free-running dual-comb systems, besides the excellent common noise suppression from the single cavity design, is the reduction of noise sources. Diode pumped solid-state lasers are highly promising in this context since their low-dispersion cavities reduce the coupling between pump relative intensity noise (RIN) and jitter, they can utilize low-noise single-mode (SM) pump diodes, and they can be passively cooled with minimal coupling to environmental noise sources. The obtainable output power of a dual-comb system pumped by one fiber-coupled SM pump diode delivering slightly below 1 W is on the 100-mW scale per comb. This is well aligned with the power requirements of modern high efficiency photoconductive antennas for THz generation or as a seed in a Master Oscillator Power Amplifier (MOPA) architecture.

*Results*

Here, we present a free running single-cavity spatially multiplexed 1.18 GHz solid-state dual-comb oscillator. The achievable high repetition rate difference combined with the low-noise performance of the laser allows for computational comb-line tracking and coherent averaging in time-domain spectroscopy (TDS) applications. We demonstrate this ability with an absorption measurement of $C_2H_2$ (acetylene) in a 20 cm long, 1 bar gas cell at the laser's wavelength around 1.05 µm. Further, the 0.85-ns delay scan range of the laser is ideally suited for high resolution THz metrology with rapid single trace update rates. We perform proof-of-principle experiments with efficient photoconductive antenna devices. In a THz spectroscopy measurement, we reach a peak spectral dynamic range of 55 dB in an integration time of 2 s, allowing to detect absorption features up to 3 THz.

The paper is structured as follows: In the first section we present the dual-comb oscillator and its noise performance. In the second section we demonstrate the TDS measurements on $C_2H_2$. Section 3 discusses the timing noise in ETS applications with adaptive sampling. The fourth section focuses on the THz-TDS and thickness measurements.

### 1. GHz dual-comb oscillator

The layout of the dual-comb oscillator is shown in Fig. 1(a). The linear confocal laser cavity is spatially multiplexed with a monolithic biprism (179° apex angle) yielding separate spots on the active elements (gain crystal and SESAM) and thereby mitigating crosstalk. Note that the actual multiplexing of the cavity is implemented in the vertical for symmetry reasons, but is shown in Fig. 1(a) in the horizontal for simplicity. The beams are separated by 1.6 mm on the high reflection (HR) coated biprism, which allows for a continuous tuning of the repetition rate difference in a range of [-175, 175] kHz by lateral translation of the biprism. Technical details of the dual-comb laser cavity are described in the methods section.

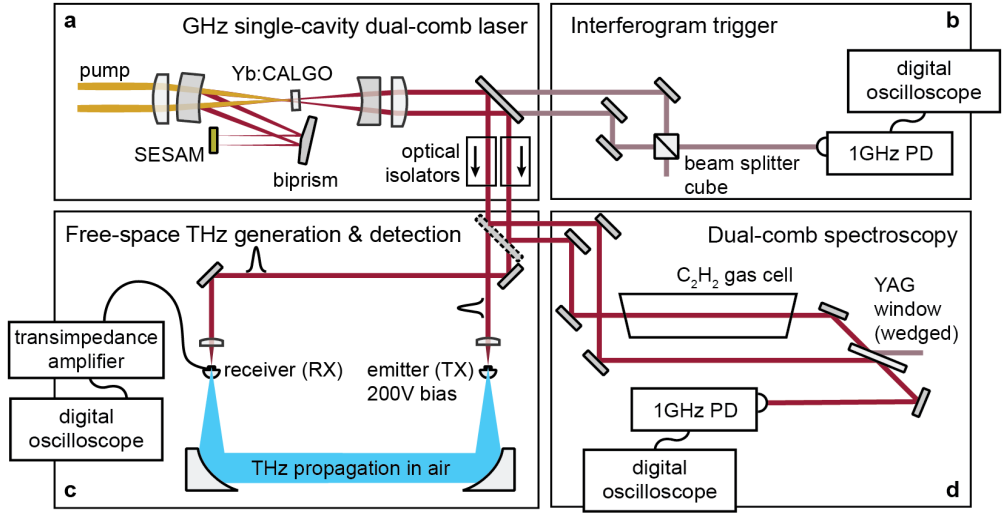

**Fig. 1:** Schematic of the setup: **(a)** Solid-state SESAM modelocked GHz dual-comb oscillator based on spatial multiplexing with a biprism in a confocal cavity arrangement, **(b)** interferogram trigger by coherent overlap of the two combs at a non-polarizing beam splitter cube, **(c)** THz time domain spectroscopy setup with efficient free-space photoconductive antennas for THz generation and detection, **(d)** dual-comb spectroscopy on acetylene ($C_2H_2$) gas cell.

### 1.1. Laser output performance

Both combs show simultaneous self-starting and robust modelocked operation over an average output power range from 80 mW to 110 mW per comb, limited by the available pump power. The combs have almost identical optical properties. The power slope is linear and the laser reaches an optical-to-optical efficiency of 23% for the highest power operation point (Fig. 2(a), sum of comb 1 and comb 2 power). The shortening of the pulse duration with increasing intracavity power follows the expected inverse scaling according to soliton formation (Fig. 2(a)). Specifically, at the highest power operation point the pulses reach a duration of 77 fs measured with second harmonic autocorrelation (Fig. 2(d)) with a full width half maximum (FWHM) of 16 nm in the optical spectrum (Fig. 2(b)) and a similar center wavelength of 1058 nm (comb 1) and 1057 nm (comb 2). We observe a clean radio frequency (RF) spectrum for both combs at a fundamental repetition rate $f_{rep} \approx 1.1796$ GHz (Fig 2(c)). The repetition rate difference is here set to $\Delta f_{rep} = 21.7$ kHz.

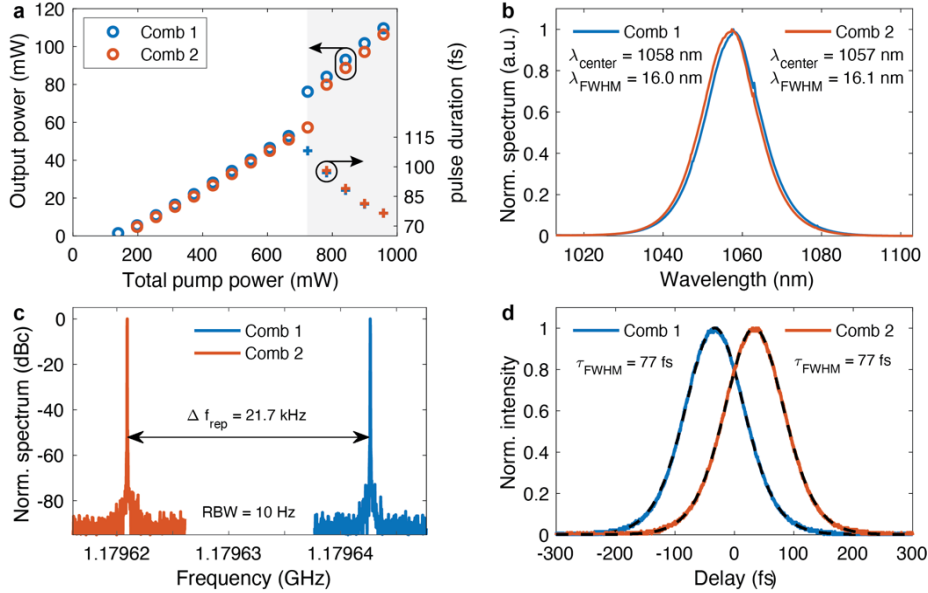

**Fig. 2:** Characterization of the dual-comb laser output with both combs operated simultaneously: **(a)** Average output power and pulse duration as a function of pump current. Detailed modelocking diagnostics are shown in (b) – (d) for the highest power operation point of the laser (as used for later measurements). **(b)** Optical spectrum. **(c)** Radio frequency spectrum of each comb around the repetition rate of the oscillator at a repetition rate difference of 21.7 kHz. **(d)** Pulse duration measurement via second harmonic autocorrelation. The pulse durations $\tau_{FWHM}$ are obtained via deconvolution assuming a sech² pulse shape (dashed lines correspond to sech² fit).

## 1.2. Dual-comb oscillator noise performance

We characterize the laser's relative intensity noise (RIN) and timing jitter. Details of these measurements are given in the supplementary material. First, we analyze the RIN of each individual comb. The root mean square (RMS) intensity noise, obtained from the integrated relative intensity noise (RIN) power spectral density (PSD), amounts to < 0.01% (integration range [10 Hz, 10 MHz]) for both combs in the free-running case (Figs. 3 (a,c)). Even lower RIN can be obtained by active stabilization of the pump power via a feedback loop. With pump stabilization (see methods for implementation details), we obtain a 15-dB suppression of the RIN for both modelocked laser outputs in the frequency range up to 100 kHz, yielding a factor 2 improvement for the integrated RMS intensity noise (Figs. 3 (a,c)) which brings it close to the ultra-low values of $3.1 \cdot 10^{-5}$ [1 Hz, 1 MHz] of our recently reported results for a multimode pumped 80 MHz oscillator [43]. Such RIN levels are beneficial for pump-probe studies, e.g. picosecond ultrasonics and time-domain thermoreflectance analysis [45].

The phase noise of the individual combs is shown in Figs. 3 (b,d). In the frequency range from 2 kHz to 100 kHz the timing jitter power spectral density (PSD) exhibits a relatively smooth drop versus frequency. When applying pump feedback, the noise in this frequency range is uniformly suppressed by about 10 dB, which indicates that noise in this band corresponds to pump RIN. The integrated timing jitter amounts to 2.4 fs and 6.4 fs (integration range [2 kHz, 1 MHz]) for the pump RIN stabilized and free running case, respectively. At lower frequencies below 2 kHz, the jitter is no longer RIN-dominated but is instead attributable to mechanical noise sources, as expected for our non-optimized optical breadboard implementation of the cavity.

A measure that is critical for any coherent averaging application with a dual-comb source is the relative timing or phase noise between the two combs, i.e. the timing noise of $\Delta f_{rep}$, as shown by the curves labeled "uncorrelated" in Figs. 3 (b,d) and determined via the method presented in [46]. This quantity (i) determines the stability of the timing axis in equivalent time sampling applications by the ratio $\Delta f_{rep}/f_{rep}$, (ii) is a dominant contributor to noise in the radio-frequency (RF) comb lines involved in coherent dual-comb spectroscopy, and (iii) reveals the extent to which the common-cavity architecture suppresses noise. Our measurements of the uncorrelated noise show that the mechanical noise sources (visible on the individual $f_{rep}$ measurements for frequencies <2 kHz) are strongly suppressed. In the free-running configuration (no pump feedback), the noise at higher frequencies is also suppressed, resulting in a high common noise suppression of approximately 20 dB throughout the full frequency range up to 100 kHz (where the noise-floor of the measurement is reached), except for one anticorrelated mechanical resonance in the system around 450 Hz. The suppression of the >2-kHz components occurs because both combs share the pump laser.

Interestingly, the pump feedback does not alter the uncorrelated noise significantly even though feedback strongly suppressed the jitter of the individual combs. Both operation regimes of the dual-comb oscillator yield <1 fs integrated uncorrelated timing jitter for the integration range [2 kHz, 1 MHz]. A possible explanation for why the uncorrelated noise is unaffected by pump RIN stabilization is the presence of asymmetric noise contributions, e.g. from the non-ideal polarization extinction ratio of the pump. Nonetheless, the noise levels with and without pump feedback are sufficiently low for our application demonstrations discussed in sections 2 and 4. Therefore, for simplicity we operate the laser in free-running mode for subsequent measurements.

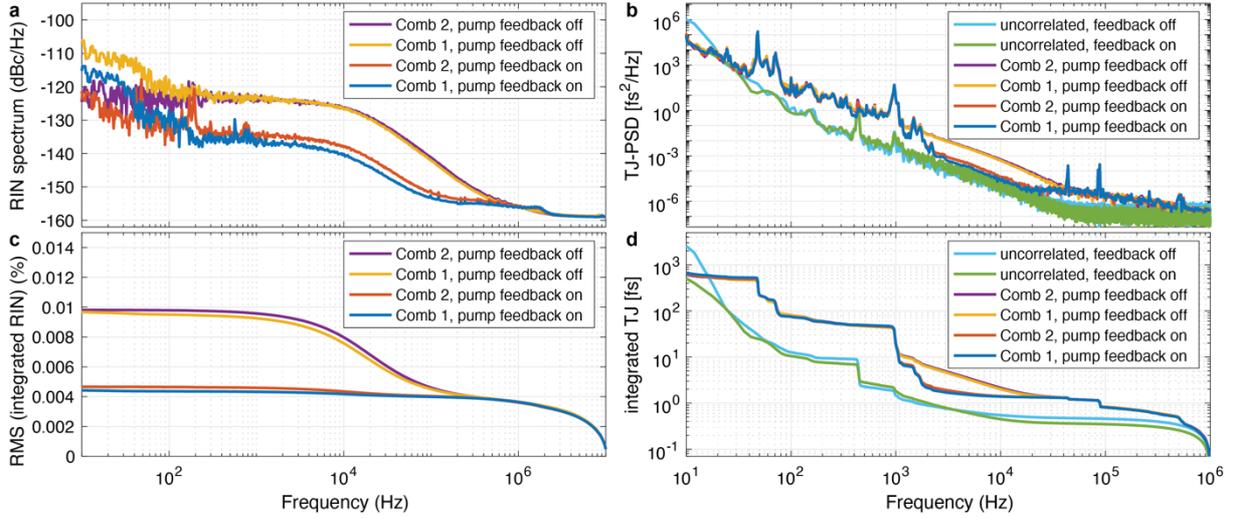

**Fig. 3: (a)** Relative intensity noise (RIN) characterization of the free running dual-comb laser with and without pump intensity stabilization (details see supplementary material) together with the RMS integrated RIN **(c)**. Both combs are operated simultaneously at the maximum output power of the laser of ≈ 110 mW per comb. **(b)** Corresponding timing jitter (TJ) characterization: One-sided power spectral density (PSD) and integrated timing jitter quantity **(d)** for the two individual combs and the uncorrelated noise part measured with the method from reference [46].

## 2. Time-domain spectroscopy on acetylene in the infrared

Based on its ultra-low noise performance, the free-running dual-comb oscillator qualifies directly for dual-comb spectroscopy (DCS). However, because of timing and other fluctuations, the interferograms formed by heterodyne beating between the two combs cannot be coherently averaged directly without a phase correction routine. The feasibility of such phase correction can be evaluated by tracking the carrier envelope phase $\Phi_{IGM}$ of the interferograms [44]. We use a relatively high value of the repetition rate difference of $\Delta f_{rep} \approx 22$ kHz in order to strongly suppresses the influence of low-frequency (<2 kHz) technical noise sources. The interferograms are obtained by combining the two co-polarized combs on a non-polarizing beam splitter cube, as shown in Fig. 1(b). Fig. 4(a) shows a typical example for the temporal evolution of the second-order finite difference $\Delta(\Delta(\Phi_{IGM}))$ of the interferogram phase. Since the fluctuations are continuously bounded by $|\Delta(\Delta(\Phi_{IGM}))| \leq \pi$, it is possible to unambiguously unwrap the phase $\Phi_{IGM}$ as a function of time [44]. In the supplementary material we describe in more detail the feasibility of this phase correction when using different values of $\Delta f_{rep}$ for the presented laser.

To confirm the suitability of the source for phase-sensitive applications like DCS, we demonstrate spectroscopy of acetylene in the ro-vibrational band around 1040 nm. The setup is sketched in Fig. 1(d): one of the two output combs is sent through a 20 cm long reference gas cell filled with acetylene (1 bar, room-temperature, Wavelength References Inc.). The light is combined with the second comb on a wedged YAG window at an angle of incidence of approx. 70° under s-polarization yielding around 40% initial intensity of each individual comb in the combined port while avoiding any etalon effects or pulse replica in the detection path. The light from the combined port is attenuated and fiber coupled and the beating of the two combs is detected on a fast photodiode (Thorlabs, DET08CFC) operated in its linear response regime.

To extract the spectroscopic information of the gas target with comb-line resolution we implement the procedure of [44]: the interferogram periods are phase-corrected, transferred to the optical domain by scaling the time axis with the comb factor $\Delta f_{rep}/f_{rep}$, and summed. Fourier transform of this coherently-averaged signal combined with a frequency shift yields the spectroscopic information in the optical frequency domain with comb line resolution. The individual optical comb lines are spaced by the repetition frequency $f_{rep}$ of the dual-comb oscillator. Fig. 4(b) shows the transmission spectrum of the acetylene gas cell for a 0.8 second integration time measurement overlayed by the prediction from HITRAN data [47]. The measured and calculated spectra are in good agreement for the entire (ro-)vibrational branch of acetylene absorption around 1040 nm. Note that for a measurement with improved signal-to-noise one could filter the optical spectrum from the laser to the region of interest and send an accordingly higher power level of light at the relevant optical frequencies onto the photodiode. Here we have used the full optical spectrum provided by the oscillator output for simplicity.

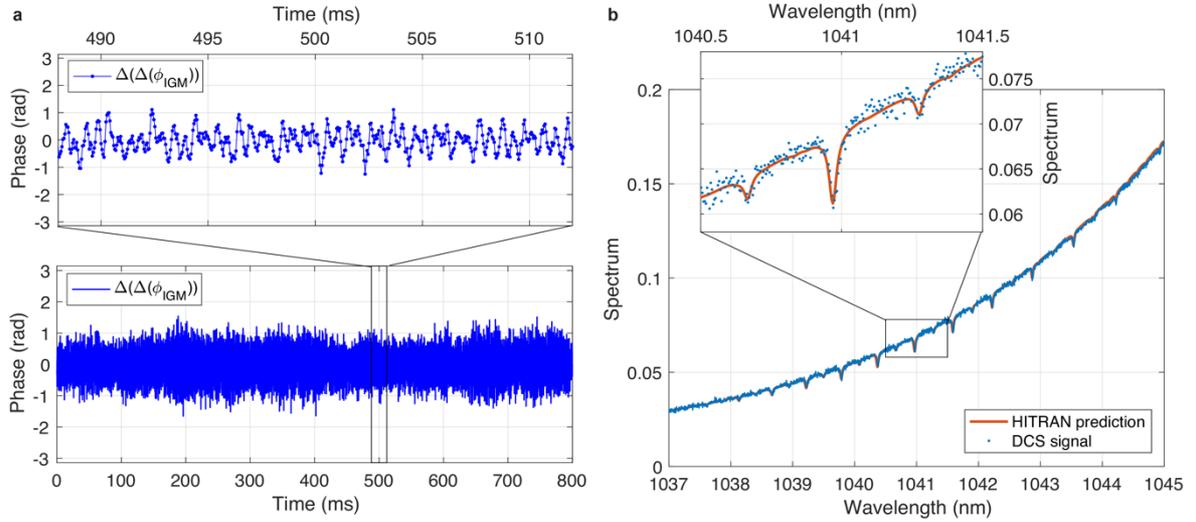

**Fig. 4: (a)** Temporal evolution of the second order finite difference $\Delta(\Delta(\Phi_{IGM}))$ of the interferogram phase $\Phi_{IGM}$ sampled at the repetition rate difference $\Delta f_{rep} \approx 22$ kHz with a zoom on the time axis. The points in the zoomed version indicate individual interferograms. **(b)** Dual-comb spectroscopy (DCS) on acetylene with the free-running GHz single-cavity dual-comb oscillator for an integration time of 0.8 seconds. Note that the absorption features from acetylene only overlap with the far wing of the optical spectrum of the laser centered at a wavelength of 1057 nm. The zoom to the absorption line around 1041 nm illustrates the spectral resolution of the DCS measurement where each point corresponds to a single comb line spaced by $f_{rep} = 1.179$ GHz in frequency or $\approx 4.3$ pm in wavelength.

### 3. Timing noise in ETS applications with adaptive sampling

In equivalent time sampling measurements, a trigger signal is generally used to avoid accumulating timing jitter over longer time scales. It is important to minimize such timing jitter because it leads to smearing out of the time axis during averaging and therefore reduces signal strength and spectral resolution. Here we use the dual-comb interferograms (IGMs) to continuously track and correct for timing drifts of the free-running oscillator. As discussed above, the IGMs are generated via heterodyne beating between the two combs (see Fig. 1(b)). IGM peaks occur whenever the pulses of the two combs overlap temporally. To determine the timing of these peaks we extract the IGM envelope with the magnitude of the Hilbert transform, and then calculate the temporal peak position by performing a second-order moment calculation. The resulting IGM peak times can be interpreted as delay zero in the context of an ETS measurement. By interpolating linearly between these peaks, we obtain the optical delay between the two pulse trains at all times during the measurement.

The accuracy of the so-obtained optical delay axis can be analyzed via the fluctuations in the time between subsequent IGM peaks, which corresponds to period jitter. Although this jitter can be obtained via the IGM peaks (and this is the approach we use for adaptive sampling in the THz-TDS measurements themselves), further insights into the timing characteristics of the laser can be obtained via the phase noise power spectral density (PN-PSD) of the underlying fluctuations in $\Delta f_{rep}(t)$, as shown in Fig. 3(b) and obtained by the method of [46].

A general approach to obtain the period jitter is via a weighted integral of the PN-PSD. For a signal described by phase $\Phi(t)$ and corresponding one-sided phase noise power spectral density $S_\Phi(f)$, the period jitter is given by [48]:

$$\tau_\Phi^{period}(\Delta f_{rep}) = \frac{1}{2\pi f_{rep}} \cdot \sqrt{\int_{f_{min}}^{f_{max}} \chi(f, \Delta f_{rep}) S_\Phi(f) df}.$$

Here $\chi(f, \Delta f_{rep}) = \left|1 - \exp\left(-2\pi i \frac{f}{\Delta f_{rep}}\right)\right|^2$ is the sampling frequency $\Delta f_{rep}$ dependent weighting factor and $f_{min}, f_{max}$ are integration limits for the offset frequency f in the PN-PSD.

In the context of ETS, the phase $\Phi(t)$ is related to the time varying repetition rate difference via $d/dt(\Phi(t)) = 2\pi \Delta f_{rep}(t)$ and the nominal period is given by $T_{sample} = 1/\langle \Delta f_{rep} \rangle$ with $\langle \cdot \rangle$ indicating the time averaged repetition rate difference. The period jitter can, however, be misleading in this context since it is influenced by slow drifts even though adaptive sampling corrects for these. To address this, we determine the part of the period jitter that adaptive sampling cannot correct for. Due to aliasing effects, the high frequency noise at $f > \Delta f_{rep}$ is partly projected to frequencies below $\Delta f_{rep}$ which is the reason for still finite contributions in the TJ-PSD at those frequencies.

Instead of performing an experiment for every setting of the repetition rate difference $\Delta f_{rep}$ we extract the information directly from the phase $\Phi(t)$ obtained with the beatnote measurement according to references [44,46]. To simulate the adaptive sampling step, we calculate the corrected phase $\Phi_{corr}(t, T_{sample}) = \Phi(t) - L_\Phi(t, T_{sample})$. Here $L_\Phi(t, T_{sample})$ is the linear interpolation of the continuous phase $\Phi$ between the grid points $t = n \cdot T_{sample}$ with $n \in \mathbb{Z}$. In Fig. 5(a) the uncorrelated timing jitter PSD is shown together with the corresponding adaptive sampling corrected PSDs for simulated repetition rate differences of 1, 5 and 22 kHz. Applying the period jitter formalism for different sampling frequencies $\Delta f_{rep} = 1/T_{sample}$ results in the curve for $\tau_{\Phi_{corr}}^{period}(\Delta f_{rep})$ presented in Fig. 5(b). For the free-running dual-comb oscillator we find an RMS timing error of the optical delay axis below 1 fs for a repetition rate detuning $\Delta f_{rep} > 18$ kHz and below 10 fs for a repetition rate detuning $\Delta f_{rep} > 1$ kHz after adaptive sampling. Note that the technical noise below 1 kHz could be mitigated in a mechanically optimized system, since the current setup was constructed on an optical breadboard with standard mirror mounts and 5-cm-high pedestals. For the THz-TDS application demonstration (discussed below) we operate the dual-comb oscillator in two configurations: at $\Delta f_{rep} = 22$ kHz where these technical noise sources are fully negligible, and $\Delta f_{rep} = 1$ kHz where the adaptive sampling period jitter of 10 fs is still small compared to the expected fastest temporal features of > 200 fs (considering a maximum THz frequency of 5 THz).

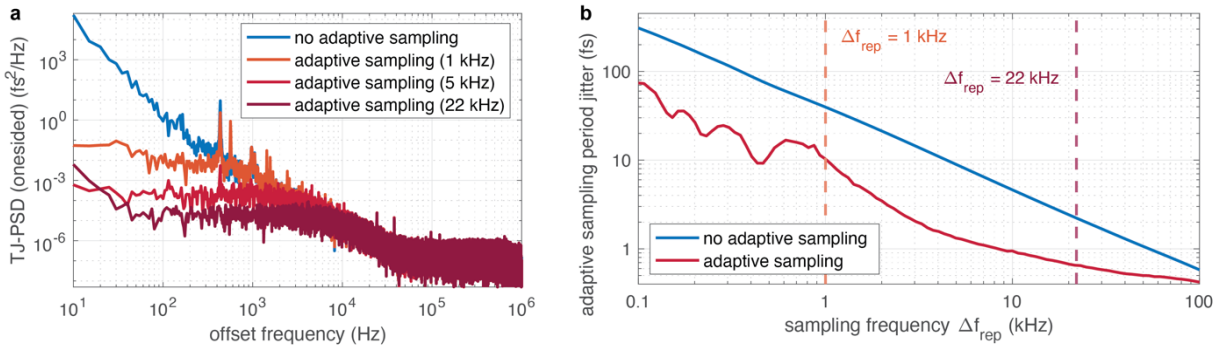

**Fig. 5: (a)** Timing jitter power spectral density (TJ-PSD) of the uncorrelated free-running dual-comb for different adaptive sampling conditions. Three different adaptive sampling cases are shown (corresponding to $\Delta f_{rep}$ values of 1 kHz, 5 kHz, and 22 kHz). **(b)** Period jitter of the optical delay axis after adaptative sampling for the free-running dual-comb oscillator under different sampling frequencies, i.e. settings of the repetition rate difference $\Delta f_{rep}$.

### 4. *Terahertz time-domain spectroscopy*

For the THz experiments we direct the light of the two combs onto two free-space photoconductive antennas (Fig. 1(c)). In the active region of the emitter device, each laser pulse generates a localized charge cloud that is accelerated in the 50 µm gap between the two electrodes by the bias electric field (40 kV/cm) and thereby generates pulsed THz radiation. The ultra-fast trapping time of the Iron-doped InGaAs material platform used for the photoconductive antenna enables short THz-pulses with frequency content up to >6 THz [49].

The generated THz radiation is collimated and refocused onto the receiver device by a pair of silicon ball lenses (mounted directly to the photoconductive antennas) and metallic off-axis parabolic mirrors. In the receiver device the optical pulses from the second comb act as a gate in order to optoelectronically sample the THz wave. More specifically, each optical pulse generates a charge cloud in the 10 µm antenna gap that is accelerated by the electric field of the THz wave, thereby inducing a small electrical current in the nA – µA range which is transimpedance amplified and detected on an oscilloscope.

To ensure no optical feedback between the THz photoconductive antennas and the laser oscillator, both free-space beam paths include a Faraday isolator (EOT, PAVOS+). The optical power in the emitter and receiver arm is controlled by a pair of half-waveplate and polarizing beam splitter. The beam is focused to a sub 50 µm spot ($1/e^2$ diameter) on the emitter with a f = 50 mm aspheric lens and a sub 10 µm spot on the receiver with a f = 20 mm lens. The positive dispersion added by the transmissive optics and the isolator crystal is compensated by negative dispersion (total around -4000 fs$^2$) from chirped mirrors to ensure compressed 77 fs pulses on the photoconductive devices.

For averaging, we use the IGM signal (described in section 3) to implement adaptive sampling of the THz time trace with linear interpolation of the optical delay axis. The result of 2 s integration or ~44000 averages is shown in Fig. 6. The main THz peak around zero optical delay repeats at $1/f_{rep} \approx 850$ ps (which marks the end of the scan window), and is followed by oscillations caused by the free induction decay of water vapor in the free space THz beam path of approximately 30 cm. The absorption features are more clearly visible in the spectral domain obtained via Fourier transformation with an apodization window of 500 ps. We apply this apodization window to suppress the influence of a feature on the THz time trace around an optical delay of 600 ps that we discuss in more detail in section 4.2. The reduced optical delay results in a spectral resolution of 2 GHz in the THz spectrum. Under these conditions we find a peak dynamic range of 35 dB in the THz power spectral density that allows to resolve absorption features up to optical frequencies of 3 THz (Fig. 6(c)). The noise floor was determined from a separate time trace where only the receiver device is illuminated by the optical frequency comb and no THz radiation is generated on the emitter side. The processing of the background trace is identical to the processing of the signal trace plus a final smoothing step in the frequency domain with a moving average.

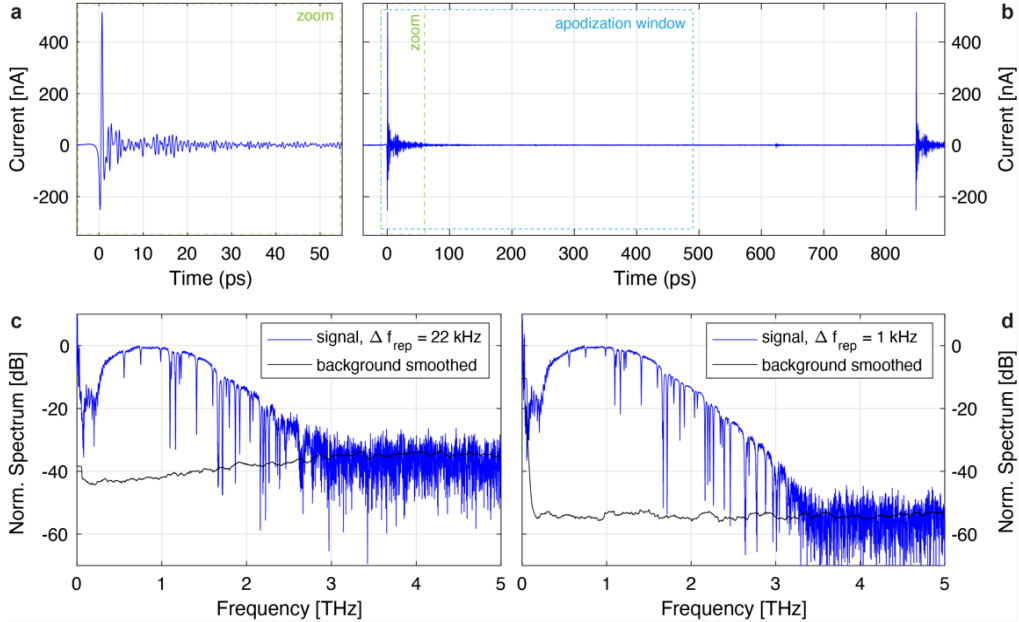

**Fig. 6: (a)** Zoom to the first 50 ps of a THz-signal time trace **(b)** obtained after 2 s integration time or ~44 k averages for the full optical delay range of $1/f_{rep} = 850$ ps at a repetition rate difference of the dual-comb laser of ~22 kHz. The bias voltage applied to the emitter has been 200 V and the average optical powers reaching the emitter and receiver have been 80 mW and 30 mW respectively. Note that a digital bandpass filter has been applied that limits the signal to THz frequencies in the range [50 GHz, 5 THz]. The first 50 ps delay range indicates clear free induction decay from absorption in the free-space THz beam path. **(c)** Power spectral density of the THz signal obtained from (b) via Fourier transformation and apodization with a 500 ps window yielding 2 GHz spectral resolution and a dynamic range of 35 dB. **(d)** THz spectrum with an increased dynamic range of 55 dB obtained from a 2 second integration time with improved amplifier noise at a lower update rate of $\Delta f_{rep} = 1$ kHz. In both cases the smoothed background is obtained from a corresponding separate time trace where the free space THz beam path was blocked. The clear absorption features steam from water absorption in the air path. Note that the absorption strength is changed due to different humidity conditions for the two measurements (late summer for (c) and early winter for (d)).

The obtained dynamic range in the THz spectrum at this high update rate ($\Delta f_{rep} \approx 22$ kHz) is largely limited by the noise-figure of the transimpedance amplifier. Operating the laser at a high repetition rate difference requires sufficient radio frequency (RF) detection bandwidth for the readout of the receiver device. The optical THz frequency $\nu_{THz}$ is mapped to the RF regime according to the equivalent time scaling factor $\Delta f_{rep}/f_{rep}$:

$$\nu_{RF} = \nu_{THz} \cdot \frac{\Delta f_{rep}}{f_{rep}}.$$

To detect THz-frequencies up to 5 THz this implies a required RF bandwidth of 93 MHz. Amplification of a weak signal with high gain-bandwidth and low noise is challenging. In our detection scheme we use a transimpedance amplifier with a 3 dB bandwidth of 200 MHz and a transimpedance gain of $10^4$ V/A (Femto HCA-S) followed by a broadband low noise voltage amplifier (Femto DUPVA-1-70) with a voltage gain of 30 dB. Lastly, we use a 200-MHz anti-aliasing filter (Minicircuits BLP-200+) before digitalization with the oscilloscope (Lecroy WavePro 254HD). A detailed discussion on the obtained dynamic range under these conditions is provided in section 4.1. To prove the claim about the amplifier limitations to the dynamic range, we performed additional measurements at an update rate of 1 kHz resulting in a relaxed requirement for the RF detection bandwidth of $\approx 4.2$ MHz (for THz frequencies up to 5 THz). At the same time, the low noise performance of the free-running dual-comb oscillator and the adaptive sampling step result in a period jitter of <10 fs (section 3). This ensures that spectroscopic information at frequencies <5 THz is not washed out in a subsequent averaging step of the time traces. The replacement of the HCA-S amplifier with a DHPCA-100 amplifier (FEMTO) (transimpedance gain $10^5$ V/A, input equivalent noise current 480 fA/$\sqrt{Hz}$, RF bandwidth 3.5 MHz) results in the expected 20 dB improvement to the signal-to-noise of the PSD of the THz signal (Fig. 6 (d)). For both configurations ($\Delta f_{rep} \approx 22$ kHz and $\Delta f_{rep} \approx 1$ kHz) the THz spectra show identical sharp absorption peaks that can be identified as water absorption. Fig. 7 shows the comparison of these absorption peaks for the $\Delta f_{rep} = 1$ kHz case with the HITRAN prediction [47]. The very good agreement for the measured position and relative strength of the absorption peaks with the HITRAN prediction indicates a well-calibrated and linear optical delay axis in our free-running dual-comb THz measurements.

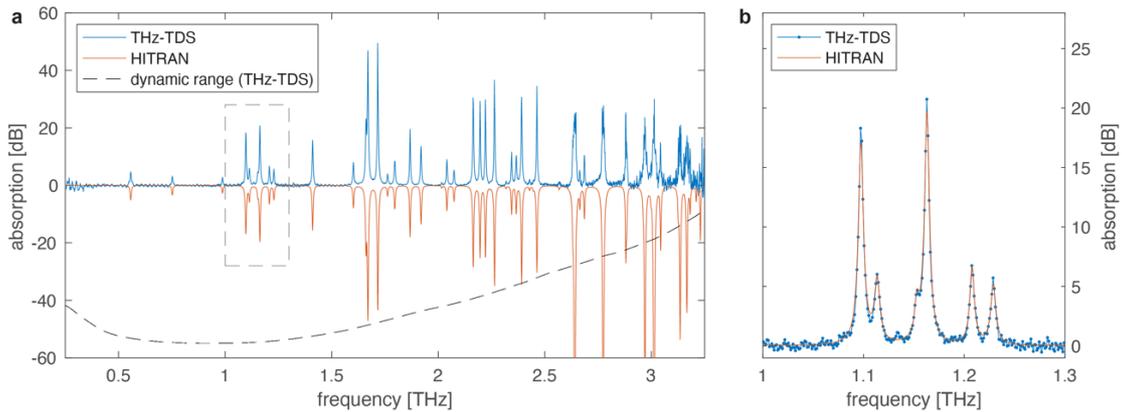

**Fig. 7:** **(a)** Comparison of the absorption features from a $\approx 30$ cm free-space air path measured via THz-TDS and the prediction from HITRAN for a water ($H_2O$) vapor concentration of 1.1 %. The THz-TDS absorption spectrum is obtained from the transmission spectrum (Fig. 6(d)) via subtraction of the THz-spectrum envelope (see appendix for details). The positions of the absorption peaks are in excellent agreement. For high frequencies, the absorption strength deviates when the predicted peak absorption strength exceeds the dynamic range of the THz-TDS measurement. **(b)** Zoom to the region between 1 THz and 1.3 THz to illustrate the spectral resolution of $\approx \mathbf{1.2}$ GHz in the THz-TDS measurement which samples well the individual absorption peaks. The THz-TDS measurement was obtained at a repetition rate difference $\Delta f_{rep} = 1$ kHz with a total integration time of 2 s.

### 4.1. Discussion of dynamic range in the THz-TDS measurements

For a direct comparison of the performance of the THz photoconductive devices in our experiment with reported literature values it is important to consider the signal strength, optical delay range, and integration time. For the devices used in our experiments, reference measurements have been performed using a laser with a driving wavelength of 1550 nm and a pulse repetition rate of 80 MHz. Under these test conditions, the obtained peak THz signal current strength amounts to 500 - 700 nA with 20 mW of optical power on both, the emitter and the receiver. Here, we explore the operation of these Fe-doped PCA devices using Yb laser technology for the first time. Despite the drastically different configuration (1050 nm wavelength and 1.2 GHz repetition rate) we obtain comparable performance in the generated THz signal current (515 – 550 nA). The average optical powers of 80 mW on the emitter and 30 mW on the receiver correspond to pulse energies that are considerably below the pulse energy damage threshold of the photoconductive devices due to the high GHz repetition rate of the laser, in contrast to the test measurements at a pulse repetition rate of 80 MHz. The required increased optical power in our experiments is explained by the photon number scaling between a 1550 nm and 1050 nm driver.

Although our signal strength is comparable to reference measurements, we obtain a significantly lower dynamic range. A high dynamic range of 105 dB in the THz power spectrum from a similar pair of photoconductive emitter and receiver devices has been reported for mechanical delay sweeps with an optical delay of 60 ps and a total integration time of 60 s [50]. In comparison, we obtained a dynamic range of 35 dB for the $\Delta f_{rep} \approx 22$ kHz configuration, and 55 dB for the $\Delta f_{rep} \approx 1$ kHz configuration. This discrepancy can be partially explained by the amount of averaging involved. We scan a longer delay range, which reduces the dynamic range (DR). To compare our results, note that the DR of a THz-TDS measurement scales like $(T_{\text{meas}}/T_{\text{range}}^2)$ for a measurement integration time $T_{\text{meas}}$ and temporal optical delay range $T_{\text{range}}$. In our apodized case, $T_{\text{range}} = 500$ ps so that the effective measurement time with the 2-s oscilloscope trace is $2\,\text{s} \cdot 500/850 = 1.18\,\text{s}$. Consequently, $(T_{\text{meas}}/T_{\text{range}}^2)$ is $\approx 3530$ times smaller (35.5 dB).

Most of the remaining discrepancy can be explained via the measurement's electronic noise floor, which is related to the transimpedance amplifier used. Mechanical delay-line based systems involve much slower sweeps of the optical delay, limiting the required detected RF frequencies to a few 10 kHz. Under these conditions, the input equivalent noise current of low-noise transimpedance amplifiers can be as low as $43\,\text{fA}/\sqrt{\text{Hz}}$ for a transimpedance gain of $10^7$ V/A, compared to $4900\,\text{fA}/\sqrt{\text{Hz}}$ for the measurement performed with $\Delta f_{rep} = 22$ kHz. The influence on dynamic range can be obtained by taking the squared ratio between the noise levels, which for the 22-kHz configuration corresponds to $(4900/43)^2 \approx 40$ dB. Accounting for this electronic factor and the temporal scaling factor, our reported dynamic range of 35 dB should theoretically scale to 35 dB + 40 dB + 35.5 dB = 110.5 dB under the conditions used in reference [50]. For the $\Delta f_{rep} = 1$ kHz configuration the experiments were performed with a transimpedance amplifier that exhibits a factor 10 better input equivalent noise current ($480\,\text{fA}/\sqrt{\text{Hz}}$) that yielded the expected 20 dB improvement to the THz power spectral density (Figs. 6 (c,d)). For this configuration we obtain a similar scaling, with 55 dB DR from the measurement, 35 dB from the temporal scaling factor, and $(4900/480)^2 = 21$ dB for the amplifier. While these calculations explain the main effects, it should be noted that the dynamic range can also be limited by the receiver antenna itself, so further improvements to the amplifier would have to be tested experimentally.

### 4.2. High-precision thickness measurements and THz pulse reflections

Next, we demonstrate the ability of the THz frontend to measure the optical and physical thickness of samples inserted into the free-space THz beam path. Here, we insert a $(2.0 \pm 0.2)$ mm thick c-cut sapphire window into the beam path. Fig. 8 shows the obtained THz time-traces versus optical delay from a single delay scan with an update rate of 1 kHz (repetition rate difference $\Delta f_{rep}$ of the laser set to 1 kHz) and after 2 s of averaging for the case with and without the additional sapphire window. Note, that the time zero is unchanged and determined by the interferogram trigger in the infrared (IR) for both cases. This allows us to identify the delays $\tau_1$ to $\tau_3$ for the main THz-pulse including etalon effects of the sapphire window around zero optical delay (as illustrated in Fig. 8(b)). Additionally, we can identify delays $\tau_4$ to $\tau_6$ at optical delays around 600 ps that correspond to propagation of the THz pulse through the free-space region between emitter and receiver of in total 3 times instead of once (Fig. 8(c)). This occurs because a small fraction of the THz light is reflected by the receiver back into the free space path, propagates back to the emitter, and is reflected again towards the receiver. The contributions to the different observed delays from the optical and physical thickness of the window are summarized in Table 1. We find the values for the physical thickness l = (2.094 ± 0.007) mm of the sapphire window and the group index $n_g$ = 3.109 ± 0.010 at an optical frequency of ~1 THz via a maximum likelihood fit to the physical model. The stated errors correspond to the 1 σ errors from the fit. Both values agree well with the mechanical thickness tolerance of the window and the group index reported in literature, respectively.

Further, the self-consistent fit with negligible uncertainty confirms the origin of the artifact in the original THz time trace without the sapphire window at around 600 ps optical delay to be the subsequent reflections of the THz waveform on the receiver and emitter device in the THz free-space path.

**Table 1:** Contributions to the optical delay of the THz waveform due to the insertion of a sapphire window into the free-space THz beam path. $n_g$ denotes the group index of sapphire along its c-axis, $l$ the physical thickness of the window and $c$ the vacuum speed of light.

|  | Optical delay contribution (ps) | Relation to sapphire window properties |
|---|---|---|
| $\tau_1$ | 14.767 | $(n_g-1) \cdot l/c$ |
| $\tau_2$ | 58.220 | $(3n_g-1) \cdot l/c$ |
| $\tau_3$ | 101.635 | $(5n_g-1) \cdot l/c$ |
| $\tau_4$ | 44.215 | $(3n_g-3) \cdot l/c$ |
| $\tau_5$ | 87.619 | $(5n_g-3) \cdot l/c$ |
| $\tau_6$ | 130.956 | $(7n_g-3) \cdot l/c$ |

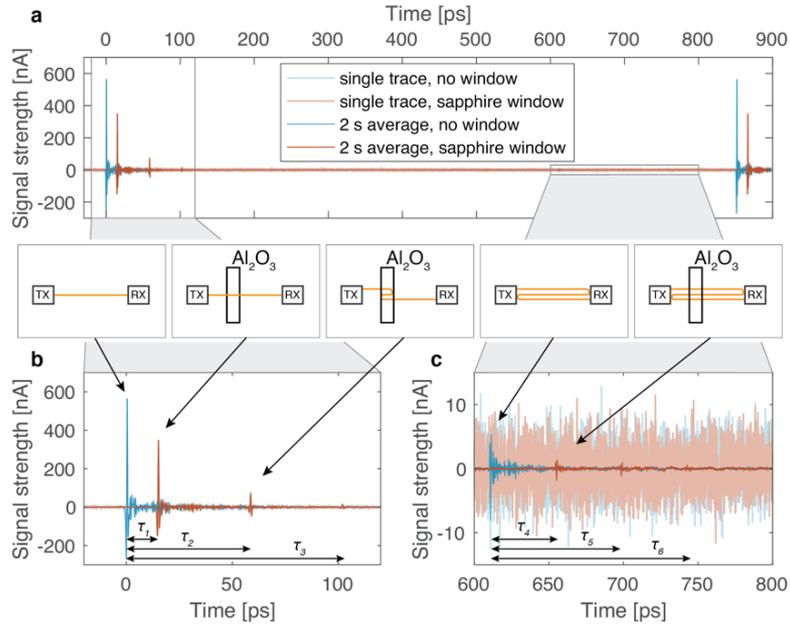

**Fig. 8:** Physical thickness and group index measurement of a 2 mm c-cut sapphire window. The window introduces an optical delay of the THz waveform between the emitter (TX) and the receiver (RX) compared to the time zero reference provided by the IR interferogram and etalon reflections (see pictograms). The stronger reflections are clearly visible on the single optical delay scan traces obtained with an update rate of 1 kHz (a). The values for the delays $\tau_1$ to $\tau_6$ indicated in (b) and (c) are provided in Table 1. Note the scale change to the signal axis in (c) for the delay range 600 ps to 750 ps to increase the visibility of the corresponding signals that only become resolved against the noise floor after averaging. For all traces, a digital bandpass filter has been applied that limits the signal to THz frequencies in the range [50 GHz, 3 THz].

## Discussion

We have demonstrated the first spatially-multiplexed single-cavity dual-comb oscillator at GHz repetition rates pumped by a spatially single-mode diode. The confocal cavity design with a biprism operated in reflective configuration allowed for a wide tunability of the repetition rate difference up to $\pm 175$ kHz and pulse durations of 77 fs with 110 mW average power per comb. The ultra-low noise performance allowed for computational comb-line tracking of the free-running oscillator output, which in turn enabled coherently averaged dual-comb spectroscopy with close to 1 GHz spectral resolution. We demonstrated these capabilities with a proof-of-principle spectroscopy experiment on an acetylene gas cell resolving all ro-vibrational absorption features around 1040 nm in excellent agreement with the predictions from HITRAN.

Further, we applied the dual-comb oscillator output directly for efficient time-domain THz experiments probing the spectroscopic features of standard air up to frequencies of 3 THz and performed precise layer thickness measurements on a sapphire window. The THz experiments benefit from the multi-kHz update rate of the full 0.85 ns delay scan. Our results show for the first time that iron-doped InGaAs based photoconductive antennas designed for optimal operation around 1550 nm can reach state-of-the-art signal strengths when driven by gigahertz 1050-nm lasers. The 55-dB dynamic range we obtained can be well explained by the THz signal strength (which is comparable to reference measurements with megahertz 1550 nm lasers), the long delay scan range (0.85 ns), and the noise of the electronic amplifier used. Furthermore, the reduced pulse energy at GHz repetition rates allows for operation at higher average power compared to conventional systems operating at repetition rates around 100 MHz. Therefore we anticipate this new platform of low-complexity single-cavity solid-state dual-comb lasers at high repetition rate with excellent free-running timing noise performance to significantly benefit high-performance THz-TDS experiments, especially when considering the repetition rate scalability towards 10 GHz [32] and the use of power-scalable Yb-doped gain-media [44].

## *Materials and Methods*

### *M1. Technical details of the dual-comb oscillator*

The laser is pumped by a 980 nm fiber Bragg grating (FBG) wavelength stabilized pump diode delivering up to 960 mW of pump power from a single-mode fiber. Part of the output power of the pump diode is split off with a polarization-maintaining 99:1 fiber splitter (Thorlabs PN980R1A1) and can be used for feedback control of the pump intensity. The main part of the power is collimated, split equally on a non-polarization dependent thin film coating beam splitter, and recombined on a D-shaped mirror. The laser is pumped through a convex-concave cavity turning mirror that has a radius of curvature (ROC) of -35 mm and a high transmission for the pump wavelength. Before this mirror, both pump beams are focused with a 50-mm focal length aspheric lens (Thorlabs AL2550M-B) onto two separate spots in the gain crystal. The crystal is a 1.5-mm-long 3% at. Ytterbium doped CALGO gain crystal (a-cut). The pump polarization is aligned along the crystal c-axis, while both combs are polarized along the a-axis of the crystal. The crystal is oriented such that this corresponds to vertical (s-) polarization in the cavity in order reduce the sensitivity of the mirror coatings to angle of incidence changes off from zero degree. The two co-polarized output beams are coupled out via the 0.8% transmission output coupler (ROC = -25 mm). Soliton formation is achieved by adding a total of -400 fs$^2$ negative group delay dispersion by the intracavity mirrors combined with the confocal cavity design which yields a high intensity in the gain crystal. Modelocking is initiated by a semiconductor saturable absorber mirror (SESAM) with a modulation depth of $\Delta R = 1.24\%$ and a fast recovery time of below 1 ps. The two output beams are collimated by a second 50 mm focal length aspheric lens (Thorlabs AL2550M-C).

For optional feedback on the pump intensity, we use the 1% port of the polarization-maintaining 99:1 fiber splitter in the pump delivery and send this power (~10 mW) onto a 70-MHz reverse biased InGaAs photodiode. The signal from the photodiode is used as the error input signal to a PI$^2$D controller (Vescent servo driver D2-125) acting on the current modulation of the laser diode driver controller. Note, that mechanical and polarization induced noise sources on the laser output are not stabilized by the feedback loop on the pump intensity.

## *Acknowledgements*

The authors acknowledge Matthias Golling for the growth of the SESAM that is used to modelock the GHz dual-comb laser and Alexander Nussbaum Lapping for assistance in careful characterization of the SESAM parameters.

The authors further acknowledge support of the technology and cleanroom facility FIRST of ETH Zurich for advanced micro- and nanotechnology.

*Funding*
Swiss National Science Foundation (40B2-0_180933, 40B1-0_203709); European Research Council (966718)

*Contributions*
B.W., C.P. and J.P. planned the dual-comb laser cavity. B.W. implemented the laser and J.P. assisted in experimental optimization steps. B.W. performed the spectroscopy experiments. B.W. and S.C. performed the noise characterization and related optimization of the laser. C.P. developed the computer program for coherent averaging and adaptive sampling of the time-domain spectroscopy and THz data. B.W., C.P. and J.P. developed the concept of the study on the THz devices and B.W. carried out the final experiments. L.L. assisted in the initial setup of the THz devices and measurement noise optimization. L.L., R.K. and B.G. provided the THz photoconductive devices and advised in their use. B.W. and C.P. carried out the data analysis and wrote the manuscript. C.P. and U.K. supervised the project. All authors discussed the results, read and approved the final manuscript.

Disclosures
The authors declare no conflicts of interest.

Data Availability Statement
Data underlying the results presented in this paper are not publicly available at this time but may be obtained from the authors upon reasonable request.

*References*
1. H. R. Telle, G. Steinmeyer, A. E. Dunlop, J. Stenger, D. H. Sutter, and U. Keller, "Carrier-envelope offset phase control: A novel concept for absolute optical frequency measurement and ultrashort pulse generation," Appl. Phys. B **69**, 327–332 (1999).
2. A. Apolonski, A. Poppe, G. Tempea, Ch. Spielmann, Th. Udem, R. Holzwarth, T. W. Hänsch, and F. Krausz, "Controlling the Phase Evolution of Few-Cycle Light Pulses," Phys. Rev. Lett. **85**, 740–743 (2000).
3. D. J. Jones, S. A. Diddams, J. K. Ranka, A. Stentz, R. S. Windeler, J. L. Hall, and S. T. Cundiff, "Carrier-Envelope Phase Control of Femtosecond Mode-Locked Lasers and Direct Optical Frequency Synthesis," Science **288**, 635–639 (2000).
4. T. Fortier and E. Baumann, "20 years of developments in optical frequency comb technology and applications," Commun. Phys. **2**, 1–16 (2019).
5. N. Picqué and T. W. Hänsch, "Frequency comb spectroscopy," Nat. Photonics **13**, 146–157 (2019).
6. S. Schiller, "Spectrometry with frequency combs," Opt. Lett. **27**, 766–768 (2002).
7. I. Coddington, N. Newbury, and W. Swann, "Dual-comb spectroscopy," Optica **3**, 414–426 (2016).
8. K. J. Weingarten, M. J. W. Rodwel, and D. M. Bloom, "Picosecond optical sampling of GaAs integrated circuits," IEEE J. Quantum Electron. **24**, 198–220 (1988).
9. P. A. Elzinga, R. J. Kneisler, F. E. Lytle, Y. Jiang, G. B. King, and N. M. Laurendeau, "Pump/probe method for fast analysis of visible spectral signatures utilizing asynchronous optical sampling," Appl. Opt. **26**, 4303–4309 (1987).
10. N. Hoghooghi, S. Xing, P. Chang, D. Lesko, A. Lind, G. Rieker, and S. Diddams, "Broadband 1-GHz mid-infrared frequency comb," Light Sci. Appl. **11**, 264 (2022).
11. O. Kara, L. Maidment, T. Gardiner, P. G. Schunemann, and D. T. Reid, "Dual-comb spectroscopy in the spectral fingerprint region using OPGaP optical parametric oscillators," Opt. Express **25**, 32713–32721 (2017).
12. C. P. Bauer, S. L. Camenzind, J. Pupeikis, B. Willenberg, C. R. Phillips, and U. Keller, "Dual-comb optical parametric oscillator in the mid-infrared based on a single free-running cavity," Opt. Express **30**, 19904–19921 (2022).

13. S. Vasilyev, A. Muraviev, D. Konnov, M. Mirov, V. Smoslki, I. Moskalev, S. Mirov, and K. Vodopyanov, "Video-rate broadband longwave IR dual-comb spectroscopy with 240,000 comb-mode resolved data points," arXiv:2210.07421 (2022).
14. D. R. Bacon, J. Madéo, and K. M. Dani, "Photoconductive emitters for pulsed terahertz generation," J. Opt. **23**, 064001 (2021).
15. Naftaly, Vieweg, and Deninger, "Industrial Applications of Terahertz Sensing: State of Play," Sensors **19**, 4203 (2019).
16. A. G. Davies, A. D. Burnett, W. Fan, E. H. Linfield, and J. E. Cunningham, "Terahertz spectroscopy of explosives and drugs," Mater. Today **11**, 18–26 (2008).
17. M. Yahyapour, A. Jahn, K. Dutzi, T. Puppe, P. Leisching, B. Schmauss, N. Vieweg, and A. Deninger, "Fastest Thickness Measurements with a Terahertz Time-Domain System Based on Electronically Controlled Optical Sampling," Appl. Sci. **9**, 1283 (2019).
18. E. Pickwell and V. P. Wallace, "Biomedical applications of terahertz technology," J. Phys. Appl. Phys. **39**, R301–R310 (2006).
19. M. van Exter, C. Fattinger, and D. Grischkowsky, "Terahertz time-domain spectroscopy of water vapor," Opt. Lett. **14**, 1128–1130 (1989).
20. R. B. Kohlhaas, S. Breuer, L. Liebermeister, S. Nellen, M. Deumer, M. Schell, M. P. Semtsiv, W. T. Masselink, and B. Globisch, "637 µW emitted terahertz power from photoconductive antennas based on rhodium doped InGaAs," Appl. Phys. Lett. **117**, 131105 (2020).
21. U. Puc, T. Bach, P. Günter, M. Zgonik, and M. Jazbinsek, "Ultra-Broadband and High-Dynamic-Range THz Time-Domain Spectroscopy System Based on Organic Crystal Emitter and Detector in Transmission and Reflection Geometry," Adv. Photonics Res. **2**, 2000098 (2021).
22. S. Mansourzadeh, T. Vogel, A. Omar, M. Shalaby, M. Cinchetti, and C. J. Saraceno, "Broadband THz-TDS with 5.6 mW average power at 540 kHz using organic crystal BNA," (2022).
23. D. Saeedkia, ed., *Handbook of Terahertz Technology for Imaging, Sensing and Communications*, Woodhead Publishing Series in Electronic and Optical Materials (Woodhead Publishing, 2013).
24. F. Tauser, C. Rausch, J. H. Posthumus, and F. Lison, "Electronically controlled optical sampling using 100 MHz repetition rate fiber lasers," in *Commercial and Biomedical Applications of Ultrafast Lasers VIII* (SPIE, 2008), Vol. 6881, pp. 139–146.
25. T. Hochrein, R. Wilk, M. Mei, R. Holzwarth, N. Krumbholz, and M. Koch, "Optical sampling by laser cavity tuning," Opt. Express **18**, 1613–1617 (2010).
26. M. Kolano, B. Gräf, S. Weber, D. Molter, and G. von Freymann, "Single-laser polarization-controlled optical sampling system for THz-TDS," Opt. Lett. **43**, 1351–1354 (2018).
27. A. Bartels, R. Cerna, C. Kistner, A. Thoma, F. Hudert, C. Janke, and T. Dekorsy, "Ultrafast time-domain spectroscopy based on high-speed asynchronous optical sampling," Rev. Sci. Instrum. **78**, 035107 (2007).
28. O. Kliebisch, D. C. Heinecke, and T. Dekorsy, "Ultrafast time-domain spectroscopy system using 10 GHz asynchronous optical sampling with 100 kHz scan rate," Opt. Express **24**, 29930–29940 (2016).
29. S. Schilt, N. Bucalovic, V. Dolgovskiy, C. Schori, M. C. Stumpf, G. D. Domenico, S. Pekarek, A. E. H. Oehler, T. Südmeyer, U. Keller, and P. Thomann, "Fully stabilized optical frequency comb with sub-radian CEO phase noise from a SESAM-modelocked 1.5-µm solid-state laser," Opt. Express **19**, 24171–24181 (2011).
30. A. Klenner, M. Golling, and U. Keller, "High peak power gigahertz Yb:CALGO laser," Opt. Express **22**, 11884–11891 (2014).
31. T. D. Shoji, W. Xie, K. L. Silverman, A. Feldman, T. Harvey, R. P. Mirin, and T. R. Schibli, "Ultra-low-noise monolithic mode-locked solid-state laser," Optica **3**, 995–998 (2016).
32. A. S. Mayer, C. R. Phillips, and U. Keller, "Watt-level 10-gigahertz solid-state laser enabled by self-defocusing nonlinearities in an aperiodically poled crystal," Nat. Commun. **8**, 1673 (2017).
33. S. Kimura, S. Tani, and Y. Kobayashi, "Kerr-lens mode locking above a 20 GHz repetition rate," Optica **6**, 532–533 (2019).
34. M. Hamrouni, F. Labaye, N. Modsching, V. J. Wittwer, and T. Südmeyer, "Efficient high-power sub-50-fs gigahertz repetition rate diode-pumped solid-state laser," Opt. Express **30**, 30012–30019 (2022).
35. H. A. Haus and A. Mecozzi, "Noise of mode-locked lasers," IEEE J. Quantum Electron. **29**, 983–996 (1993).

36. R. Paschotta, A. Schlatter, S. C. Zeller, H. R. Telle, and U. Keller, "Optical phase noise and carrier-envelope offset noise of mode-locked lasers," Appl. Phys. B **82**, 265–273 (2006).
37. S. M. Link, A. Klenner, M. Mangold, C. A. Zaugg, M. Golling, B. W. Tilma, and U. Keller, "Dual-comb modelocked laser," Opt. Express **23**, 5521–5531 (2015).
38. T. Ideguchi, T. Nakamura, Y. Kobayashi, and K. Goda, "Kerr-lens mode-locked bidirectional dual-comb ring laser for broadband dual-comb spectroscopy," Optica **3**, 748–753 (2016).
39. S. Mehravar, R. A. Norwood, N. Peyghambarian, and K. Kieu, "Real-time dual-comb spectroscopy with a free-running bidirectionally mode-locked fiber laser," Appl. Phys. Lett. **108**, 231104 (2016).
40. B. Willenberg, B. Willenberg, J. Pupeikis, J. Pupeikis, L. M. Krüger, F. Koch, C. R. Phillips, and U. Keller, "Femtosecond dual-comb Yb:CaF$_2$ laser from a single free-running polarization-multiplexed cavity for optical sampling applications," Opt. Express **28**, 30275–30288 (2020).
41. J. Pupeikis, B. Willenberg, F. Bruno, M. Hettich, A. Nussbaum-Lapping, M. Golling, C. P. Bauer, S. L. Camenzind, A. Benayad, P. Camy, B. Audoin, C. R. Phillips, and U. Keller, "Picosecond ultrasonics with a free-running dual-comb laser," Opt. Express **29**, 35735–35754 (2021).
42. S. L. Camenzind, T. Sevim, B. Willenberg, J. Pupeikis, A. Nussbaum-Lapping, C. R. Phillips, and U. Keller, "Free-running Yb:KYW dual-comb oscillator in a MOPA architecture," Opt. Express **31**, 6633–6648 (2023).
43. J. Pupeikis, B. Willenberg, S. L. Camenzind, A. Benayad, P. Camy, C. R. Phillips, and U. Keller, "Spatially multiplexed single-cavity dual-comb laser," Optica **9**, 713–716 (2022).
44. C. R. Phillips, B. Willenberg, A. Nussbaum-Lapping, F. Callegari, S. L. Camenzind, J. Pupeikis, and U. Keller, "Coherently averaged dual-comb spectroscopy with a low-noise and high-power free-running gigahertz dual-comb laser," Opt. Express **31**, 7103–7119 (2023).
45. J. Pupeikis, W. Hu, B. Willenberg, M. Mehendale, G. A. Antonelli, C. R. Phillips, and U. Keller, "Efficient pump-probe sampling with a single-cavity dual-comb laser: Application in ultrafast photoacoustics," Photoacoustics **29**, 100439 (2023).
46. S. L. Camenzind, D. Koenen, B. Willenberg, J. Pupeikis, C. R. Phillips, and U. Keller, "Timing jitter characterization of free-running dual-comb laser with sub-attosecond resolution using optical heterodyne detection," Opt. Express **30**, 5075–5094 (2022).
47. R. V. Kochanov, I. E. Gordon, L. S. Rothman, P. Wcisło, C. Hill, and J. S. Wilzewski, "HITRAN Application Programming Interface (HAPI): A comprehensive approach to working with spectroscopic data," J. Quant. Spectrosc. Radiat. Transf. **177**, 15–30 (2016).
48. S. Meninger, "Phase Noise and Jitter," in *Clocking in Modern VLSI Systems*, T. Xanthopoulos, ed., Integrated Circuits and Systems (Springer US, 2009), pp. 139–181.
49. B. Globisch, R. J. B. Dietz, R. B. Kohlhaas, T. Göbel, M. Schell, D. Alcer, M. Semtsiv, and W. T. Masselink, "Iron doped InGaAs: Competitive THz emitters and detectors fabricated from the same photoconductor," J. Appl. Phys. **121**, 053102 (2017).
50. R. B. Kohlhaas, S. Breuer, S. Nellen, L. Liebermeister, M. Schell, M. P. Semtsiv, W. T. Masselink, and B. Globisch, "Photoconductive terahertz detectors with 105 dB peak dynamic range made of rhodium doped InGaAs," Appl. Phys. Lett. **114**, 221103 (2019).
51. G. D. Domenico, S. Schilt, and P. Thomann, "Simple approach to the relation between laser frequency noise and laser line shape," Appl. Opt. **49**, 4801–4807 (2010).

*Supplementary Material*

*S1. Noise characterization of individual combs: RIN with frequency stitching and phase noise measurements*

The RIN of the laser outputs is measured on a 70 MHz bandwidth reverse biased InGaAs photodiode. For increased sensitivity, we separate the measurement into two frequency bands (<200 kHz and >= 200 kHz) and stitch them afterwards as described in [43] For the lower frequency band up to 200 kHz, we amplify the photodiode signal with a transimpedance amplifier (Femto DHPCA-100). For the high frequency band, we split the signal from the photodiode into its AC and DC part with a bias tee (SHF BT45R), apply a 50 Ohm load on the DC part and amplify the AC part with a low noise voltage amplifier (Femto DUPVA-1-70). Both amplifiers show a flat gain in the overlapping frequency region around 200 kHz, where we stich the two measurements, and are noise optimized for the respective frequency ranges.

The phase noise of the individual combs is extracted from the 13$^{th}$ harmonic at 15.3 GHz. We send approximately 1 mW of optical power onto a fast and highly linear photodiode (Lab Buddy HLPD), bandpass filter the signal with a tunable filter (Micro Lambda Wireless MLFP-22026), amplify by 25 dB and detect it on a signal source analyzer (Keysight E5052B).

*S2. RF-comb line broadening*

The linewidth of a RF comb line from the heterodyne beating of the two combs is a measure for the relative stability of the two optical frequency combs emitted by the dual-comb laser source. As long as the width of the RF comb lines stays well below their spacing, given by the repetition rate difference $\Delta f_{rep}$, comb lines are directly resolvable via Fourier transform of the interferogram's time trace and without additional phase correction. The linewidth can be calculated according to the β-separation line formalism from the frequency noise PSD (FN-PSD) of the RF comb line. For the free-running dual-comb oscillator we calculate the FN-PSD from the uncorrelated phase noise PSD via scaling by frequency. The uncorrelated phase noise PSD itself is obtained with the method described in [46] based on heterodyne beating of the two combs with two continuous wave lasers. For the presented dual-comb laser the FN-PSD stays below the β-separation line (defined by $S_\beta(f) = (8\ln(2)/\pi^2) \cdot f$) [51]) for frequencies above 2 kHz (Fig. 9(a)). The FN-PSD crosses the β-separation line at 1.75 kHz, indicating that technical noise sources are the dominant contribution to the RF-comb linewidth.

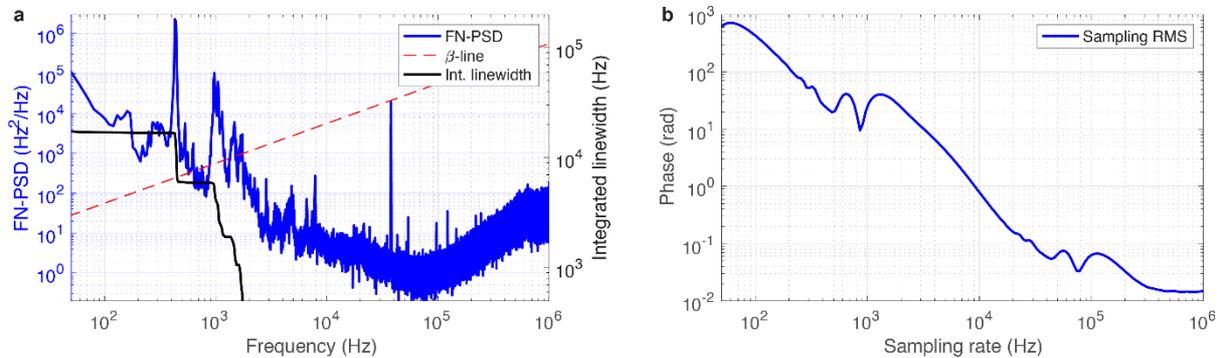

**Fig. 9: (a)** Frequency noise PSD for a single RF comb line corresponding to the center of the optical spectrum. The dashed red curve indicates the $\beta$-separation line. According to the $\beta$-separation line formalism, only noise frequencies for which the FN-PSD is above the $\beta$-separation line contribute to the broadening of the RF comb lines. Consequently, the RF-comb lines are only broadened by noise at frequencies below 1.75 kHz, i.e. integration times exceeding 500 μs. **(b)** RMS phase fluctuations ($RMS[\Delta_\tau^{(2)}(\Phi)]$) of the free-running dual-comb oscillator as a function of the sampling frequency $1/\tau$.

*S3. Limitations to interferogram phase correction for coherent averaging*

Coherent averaging for the full optical spectrum, which in the limit of infinitely long averaging time yields infinitely sharp comb lines, breaks down for more constrained noise conditions than the ones defined by the β-separation line formalism described above. Consider the case where the phase correction in the post-processing step is restricted to the phase information $\Phi_{IGM}$ directly obtainable from the interferograms. In this case, one can capture the limit of this correction by the RMS of the second order finite difference of the interferogram phase $\Delta(\Delta(\Phi_{IGM}))$ as described in the main text and in reference [44]. To generalize this information to different sampling rates $1/\tau \neq \Delta f_{rep}$, we can calculate the sampling RMS from the continuous phase $\Phi$ obtained via the heterodyne beating with the two cw lasers. Once the value for $RMS \gg 1$, it will become difficult to detect phase slips of multiples of $2\pi$ between subsequent interferograms and to unambiguously unwrap the required phase. The corresponding curve is presented in Fig. 9(c) for the free-running dual-comb. It indicates that the phase correction routine for coherent averaging will be effective provided that $\Delta f_{rep} > 10$ kHz.

## S4. THz spectrum envelope extraction

To infer the absorption spectrum from the measured THz spectrum we need to subtract the measured spectrum from its envelope. Since purging of the free-space THz path to 0% relative humidity to obtain a reference spectrum is experimentally challenging, we instead extract the envelope of the THz spectrum from the THz time trace itself. The envelope of the spectrum is governed by the impulse response of the THz emitter to an optical pulse. In cases where the free induction decay of the probed gas volume would be well separated from this main THz pulse, an apodization window around zero optical delay would yield the envelope of the spectrum. In the case of strong water absorption in air, the main THz pulse and the free induction decay signal are not separable in the time-domain. In order to still extract the main THz pulse, we construct a causal spectral filter from the HITRAN prediction for the transmission spectrum of the air path with water absorption. Dividing the measured spectrum by the predicted spectral filter yields a separable main THz peak around zero optical delay (Fig. 10). To avoid numerical instabilities, we restrict the correction to THz frequencies below 3.5 THz and HITRAN predicted absorption features with a strength of >1%. We temporally apodize the resulting signal (temporal window of 10 ps) and Fourier transform the resulting THz pulse to obtain the envelope of the THz spectrum. Note that this approach at the same time filters out temporally far distant reflections and etalon effects in the THz beam path.

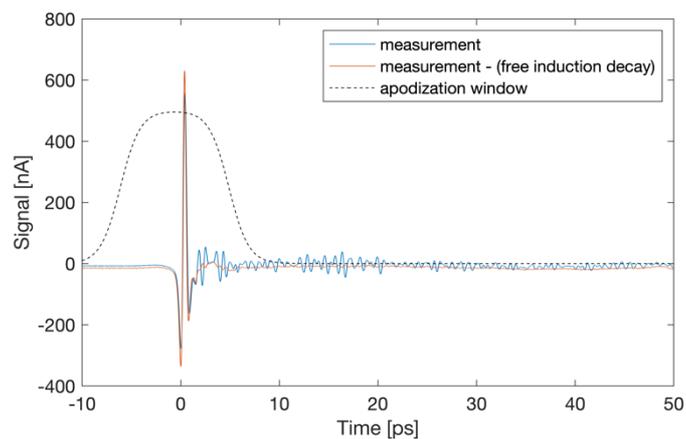

**Fig. 10:** Illustration of the THz peak extraction from the superimposed free induction decay signal of humid air.